\documentclass{aa}
\usepackage{ffbib}
\usepackage{psfig}
\usepackage{epsfig}
\def\gsim{\;\lower4pt\hbox{${\buildrel\displaystyle >\over\sim}$}\,}
\def\lsim{\;\lower4pt\hbox{${\buildrel\displaystyle <\over\sim}$}\,}

\def \sax {{\em BeppoSAX}}
\def \ros {{\em ROSAT}}
\def \src {G318.2+0.1}
\def \degmark{^\circ}

\def \hcm {\hbox {\ifmmode $ atom cm$^{-2}\else atom cm$^{-2}$\fi}}
\def \arcmin {\hbox{$^\prime$}}
\def \arcsec {\hbox{$^{\prime\prime}$}}

\def\approxgt{\mathrel{\hbox{\rlap{\lower.55ex \hbox {$\sim$}}
        \kern-.3em \raise.4ex \hbox{$>$}}}}
\def\approxlt{\mathrel{\hbox{\rlap{\lower.55ex \hbox {$\sim$}}
        \kern-.3em \raise.4ex \hbox{$<$}}}}
\newcommand{\mc}{\multicolumn}

\begin{document}


\title{X-ray emission in the direction of the SNR G318.2+0.1}


      \author{F. Bocchino\inst{1,2}
         \and A.N. Parmar\inst{1}
         \and S. Mereghetti \inst{3}
         \and M. Orlandini\inst{4}
         \and A. Santangelo\inst{5}
         \and L. Angelini\inst{6}
}
\offprints{F. Bocchino (fbocchin@estec.esa.nl)}

\institute{
       Astrophysics Division, Space Science Dept. of ESA, ESTEC,
              Postbus 299, 2200 AG Noordwijk, The Netherlands
\and
       Osservatorio Astronomico di Palermo, Piazza del Parlamento 1,
       90134 Palermo, Italy
\and   
       Istituto di Fisica Cosmica e Tecnologie Relative, CNR, Via Bassini 15, 
       20133 Milano, Italy
\and
       Istituto Tecnologie e Studio Radiazioni Extraterrestri, CNR, 
       Via Gobetti 101, 40129 Bologna, Italy
\and
       Istituto di Fisica Cosmica ed Applicazioni all'Informatica, CNR, Via 
       U. La Malfa 153, 90146 Palermo, Italy
\and
       Laboratory for High Energy Astrophysics, Code 660.2, 
       NASA/Goddard Space Flight Center, MD 20771, USA
}

\date{Received 3 Jul 2000; Accepted 17 Nov 2000}

\abstract{
We report the discovery of three X-ray sources within the radio shell
\src, one of which may be extended.  Two of the sources were detected
during the \sax\  Galactic Plane Survey and one was found in archival
\ros\  data. The fainter \sax\  source is coincident with an
ultra-compact galactic H~{\sc ii} region, and we discuss the
possibility that it can be a flaring young stellar object,
while the other \sax\  source has no obvious counterpart. The PSPC
source is consistent with emission from a foreground star.  The hard
spectrum of the brighter \sax\  source is consistent with a
non-thermal origin, although a thermal nature cannot be formally
excluded.  If this source is associated with \src, then its hard
spectrum suggests that it may be site of non-thermal electron
acceleration.
}

\maketitle

\markboth{F. Bocchino et al.}
{X-rays from G318.2+0.1}

\section{Introduction}

\src\ is a large ($40^\prime\times 35^\prime$) radio shell discovered
by \cite*{wg96} in the MOST 843~MHz survey.  The remnant was
identified by means of the criterium of the ratio of 60 $\mu$m to
radio flux density, but \cite*{wg96} have also reported of additional
polarization measurement which seems to confirm its nature. This object
is mainly composed of two non-thermal filaments at the North-West and
South-East which form two sections of the shell, plus a central region
and a bright clump to the South-West, which are thermal and correspond
to two H~{\sc ii} regions. Moreover, the ultra-compact (UC) H~{\sc ii}
region IRAS14498--5856, reported in \cite*{whr97}, is also located
towards the radio shell; this UC H~{\sc ii} region is a methanol, OH and
H$_2$O maser (GAL318.05+0.08, \cite{gm93}).  As many other newly
discovered radio shells, very little is known about \src. In particular,
we do not know its age and its distance, and the relation between the
SNR and H~{\sc ii} region detected in radio is still to be studied.

The X-ray band offers a unique opportunity to confirm the detection of
newly discovered radio SNRs. In fact, SNRs may be strong X-ray emitters
since their initial evolutionary stage, and several SNRs have been
discovered, or rediscovered, in large area surveys or in pointed
observations by X-ray satellites (see e.g., \cite{abp91,pa96}). The
detection of radio SNRs in other bands is important in order to
properly assess the nature of these objects, as well as to provide a
secure identification in cases where this is uncertain. Among other
things, X-ray studies may provide information about the distance, which
is a key parameter for SNRs, measured with difficulty in the radio
band.

In this {\it paper} we report the discovery of three X-ray sources
within the radio shell \src. Two are sources discovered during the
\sax\  Galactic Plane Survey (GPS), one of which is positional
coincident with the UC H~{\sc ii} region, and one was found in archival
\ros\  data.  In Sect.~2 we describe the observations, in Sect.~3 we
introduce the detected sources and in Sect. 4 we discuss their possible
nature and relation to \src.

\section{Observations}

The \sax\  GPS consisted of 14 pointed observations of a region of
the galactic plane between approximately ${\rm l = 310\degmark}$ to
${\rm l = 320\degmark}$ within about $\pm 1\degmark$ from the plane. Due to a
lack of suitable guide stars the region between  ${\rm l = 312\degmark}$
to ${\rm l = 316\degmark}$ could not be observed. The prime instruments for
the survey were the Medium-Energy Concentrator Spectrometer (MECS;
1.8--10~keV; \cite{bbp97}) and the Low-Energy Concentrator Spectrometer
(LECS; 0.1--10~keV; \cite{pmb97}).  The MECS consists of
two grazing incidence telescopes with imaging gas scintillation
proportional counters in their focal planes. The LECS uses an identical
concentrator system as the MECS, but utilizes an ultra-thin entrance
window and a driftless configuration to extend the low-energy response
to 0.1~keV.  The fields of view (FOV) of the LECS and MECS are circular
with diameters of 37\arcmin\ and 56\arcmin, respectively.  In the
overlapping energy range, the position resolution of both instruments
is similar and corresponds to 90\% encircled energy within a radius of
2\farcm5 at 1.5~keV. In addition, the \sax\  payload includes two
high energy instruments - the High Pressure Gas Scintillation
Proportional Counter (HPGSPC; 5--120~keV; \cite{mgs97})
and the Phoswich Detection System (PDS; 15--300~keV;
\cite{fcd97}).  The non-imaging HPGSPC consists of a single unit with a
collimator that remained on-source during the GPS observations. The
non-imaging PDS consists of four independent units arranged in pairs
each having a separate collimator. Each collimator was alternatively
rocked on- and 210\arcmin\ off-source every 96~s during the
observations.

\begin{figure}
  \centerline{\psfig{file=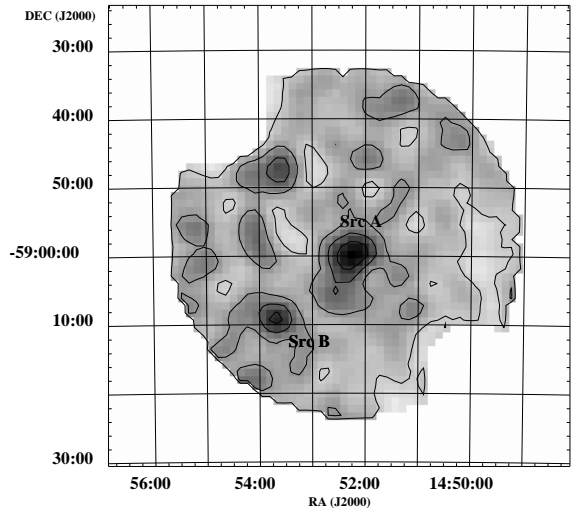,width=9cm}}
  \caption{Background subtracted and vignetting corrected 2--10~keV
  MECS image of the \sax\  GPS observation which included part of the radio
  shell \src. Contour
  levels represent 20\%, 40\%, 60\%, and 80\% of the peak intensity
   of 0.21 cts ks$^{-1}$ arcmin$^{-2}$}
  \label{mecs}
\end{figure}

The region of sky containing \src\ was observed between 1999 July 21
16:58 and July 22 06:56~UTC.  The (J2000) pointing direction was
R.A.=$14^{\rm h}\; 53^{\rm m}\; 41\fs7$,
Dec=$-58\degmark\ 59\arcmin\ 24\arcsec$.  In order to avoid solar
scattered emission and other contaminating effects, data were screened
retaining intervals when the elevation angle above the Earth's limb was
$>4^{\circ}$ and when the instrument configurations were nominal.
The screened exposures
in the LECS and MECS are 10.9~ks and 25.9~ks, respectively.  The MECS
0.5-10 keV image is shown in Fig. \ref{mecs}. The image was smoothed
using a Gaussian filter with a $\sigma$ of 1\farcm5.

Inspection of PDS data after having subtracted a contaminating source
present in one of the offset collimator positions does not reveal the
presence of any excess signal above background.  The 3$\sigma$
confidence upper limit to any 15--30~keV count rate is $<0.17$~cts~s$^{-1}$.
For a Crab-like spectrum this corresponds to $<$2~mCrab. As for the
HPGSPC, we have derived a 3$\sigma$ confidence upper limit of 0.25
s$^{-1}$ in the 10--20 keV band, which correspond to 3.5 mCrab.

In addition, we have used two \ros\  Position Sensitive Proportional
Counter (PSPC) archival observations (RP500096 and RP500192 with
exposures of 1.3 and 4.5~ks, respectively) both pointed at
R.A.=$14^{\rm h}\; 58^{\rm m}\; 19\fs2$,
Dec=$-58\degmark\ 27\arcmin\ 00\arcsec$ (J2000) which partially
overlaps the radio shell of \src. The PSPC, which covers the
0.1--2.4~keV energy range, has a circular FOV of $\sim$$2^\circ$ and a
spatial resolution of $\sim$$30^{\prime\prime}$ full width half-maximum
(FWHM) on-axis and $\sim$2\farcm5 at $45^\prime$ off-axis, near the
location of one of the source discussed below. The PSPC spectral resolution is
$E/\Delta E \sim 2$ at 1 keV. The FOVs of the \sax\  and \ros\ 
observations used here do not overlap.

\section{X-ray sources inside \src}

\begin{figure}
  \centerline{\psfig{file=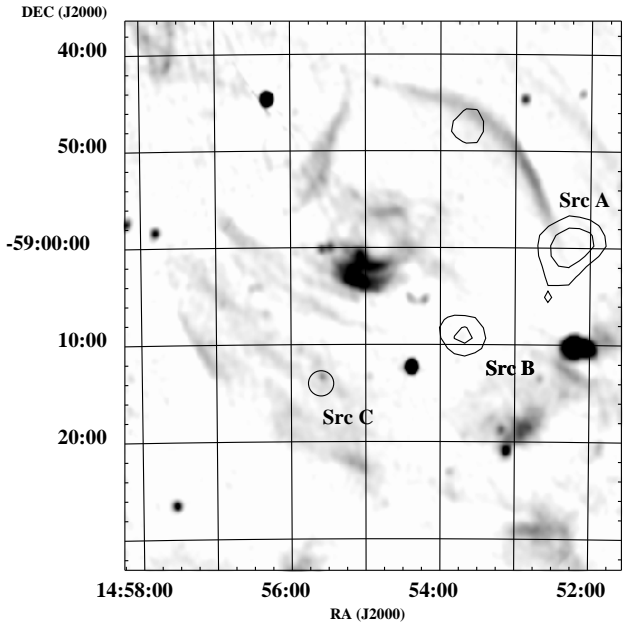,width=9cm}}
  \caption{843 MHz MOST survey image of the radio shell \src\  (from
  \protect\cite{wg96}). The 60\%, and 80\% of the peak contour
  levels of the \sax\  image which mark the locations of sources A
  and B are overlaid. A 1\farcm3 radius circle marks the
  position of PSPC source C with its uncertainty}
  \label{wg} 
\end{figure}

\subsection{Source detection}

\begin{table*}
\caption[]{List of X-ray sources towards \src. The count rate is given
in the energy range 2--10~keV (MECS) or 0.2--2.0~keV (PSPC).
Vignetting correction and background subtraction have been applied.
Position uncertainty is given as the width of 90\% confidence range}

\medskip
\begin{tabular}{lcccccl} 
\hline\noalign{\smallskip}
Source & Inst & R.A. & Dec & Uncertainty & Count rate & Significance \\
       &      & \mc{2}{c}{(J2000)} & radius (\arcmin) &(s$^{-1}$) & \\ 
\noalign{\smallskip\hrule\smallskip}
A & MECS & $14^{\rm h}$ $52^{\rm m}$ $12.4^{\rm s}$ 
  & $-58^{\rm d}$ $59^\prime$ $40\arcsec$&  1 & $(1.05\pm 0.11) \times 10^{-2}$ 
& $6\sigma$ \\

B & MECS & $14^{\rm h}$ $53^{\rm m}$ $39.5^{\rm s}$ 
& $-59^{\rm d}$ $08^\prime$ $52\arcsec$ & 1
& $(8.2\pm 1.6)\times 10^{-3}$ & $3\sigma$ \\


C & PSPC & $14^{\rm h}$ $55^{\rm m}$ $35.5^{\rm s}$ & 
$-59^{\rm d}$ $12^\prime$ $42\arcsec$ & 1.3 & 
$0.115\pm 0.008$ & $27\sigma$ \\ 
\noalign{\smallskip}
\hline
\end{tabular}
\label{xraysrc}
\end{table*}

Inside the radio boundary of \src, we have detected the three X-ray
sources which are listed in Table~\ref{xraysrc} together with their
positions, count rates and significances.  In Fig. \ref{wg}, we
reproduce the \cite*{wg96} radio map of \src\  with the position of the
X-ray sources superimposed. Sources A and B
were detected using the sliding-cell method of the {\sc ximage}
software package (\cite{gaj92}) on the 32\arcsec\ pixel \sax\  MECS
image. Some fainter X-ray emission is also present in the North-Western
part of the radio shell but it is not detected above the $3\sigma$
level and we will not discuss it further.  Source A is the brightest and is
located at the rim of the shell, between the North-West shell filament
and the bright thermal radio clump at $14^{\rm h}$ $52^{\rm m}$
$09^{\rm s}$, $-59\degmark$ $10^\prime$ discussed in \cite*{wg96}.  The
80\% (i.e.,  the innermost), 60\% and 40\% contour levels around
Source~A show some evidence of source elongation from the North-West to
the South-East which is not shown in the corresponding image of Cyg
X-1, a bright point source used as a template for the instrumental
Point Spread Function (PSF). We therefore suggest that Source~A may be
extended.  Source B is positioned roughly halfway between the shell and
the core, with no obvious radio counterpart at 843 MHz, and it is
consistent with being point-like.  No sources were detected in the LECS
image, with a $3\sigma$ upper limit of $2.7\times 10^{-3}$ cts s$^{-1}$
in a $8^\prime$ radius circle. This may be due to the much lower LECS
exposure time and/or to high X-ray absorption, which supress X-rays
below 2 keV, the only energy band where the LECS is more sensitive than
the MECS.

Source C was detected in the \ros\  All-Sky Survey (RASS, 1RXS
J145540.4-591320, \cite{vab99}) and it is also present in the WGACAT
(1WGA J1455.5-5912 and 1WGA J1455.7-5912, \cite{wga94}). The count
rates measured in different \ros\  observations are $0.11\pm 0.02$,
$0.11\pm 0.01$ and $0.115\pm 0.008$ cts s$^{-1}$ in the RASS (October
1990), RP500192 (March 1992) and RP500096 (February 1993),
respectively.  The source has a shape severely distorted by mirror
blurring since it is located at $\sim$$50^\prime$ off-axis, but its
extension is compatible with the size of the PSPC PSF at that off-axis
angle.

In order to investigate the nature of the detected X-ray source (and
therefore the relation with \src), we have performed source
cross-identifications with catalogues using SIMBAD and a search radius
of $5^\prime$. Sources A and B are coincident with very weak IRAS
sources at 12~${\rm \mu m}$. However, this band is heavily confused
by the galactic background. Source~B is the only one which has a strong
counterpart at 25, 60 and 100~${\rm \mu m}$ and is also coincident with
the UC H~{\sc ii} region IRAS14498--5856 reported by \cite*{whr97}.
As for Source~C, the star GSC08693-00036, with B$=11.4$ and V$=10.7$
falls inside its error circle, at $1^\prime$ offset.  We also inspected
the Digital Sky Survey image within $2^\prime$ from the positions of
the sources and we did not find any other obvious counterparts (except
those already found with SIMBAD).

\subsection{Spectral analysis}

\begin{figure}
  \centerline{\psfig{file=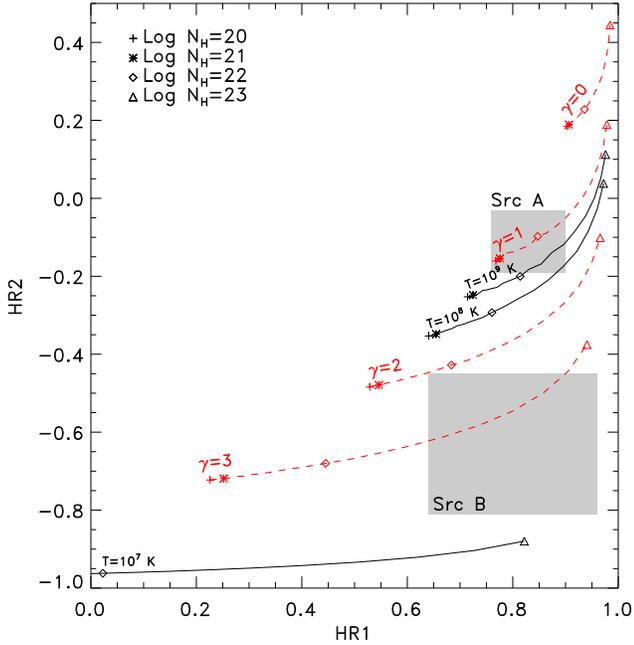,width=9cm}}
  \caption{Observed Source A and Source B MECS hardness ratios with
  uncertainties (shaded areas) and values expected from a power-law
  (gray and dashed lines) and {\sc mekal} (solid lines) emission model with different
  parameters values (derived by simulations). The lines are constant
  temperature or power-law index loci, while symbols represent $N_{rm H}$
  values}

  \label{hr_sc} 
\end{figure}

Since the sources are weak, the analysis of their hardness ratios is
a very good method to constrain the X-ray emission
process. For the MECS sources, we defined two hardness ratios in
the following way:

\begin{equation}
H\!R1=\frac{H_1-S_1}{H_1+S_1};~~~
H\!R2=\frac{H_2-S_2}{H_2+S_2}
\end{equation}

where $S_1$ is the number of counts in the 1.0--2.0 keV band, $H_1$ in
the 2.0--10 keV band, $S_2$ in the 2.0--5.0 keV band and $H_2$ 5.0--10
keV band. The counts are collected in circular area with $4^\prime$ and
$2^\prime$ radius for Source A and B, respectively, and they were
background subtracted using the MECS standard background maps, because
we have verified that the background collected in the short exposure
observations of \src\  is statistically poor.  Errors in the H and S
quantities are computed propagating the Poissonian errors.  The
uncertainties on the ratios are computed using the maximum and
minimum values that the denominators and the numerators can assume
considering their respective uncertainties. The values of $H\!R1$ and
$H\!R2$ for Source A and Source B and their uncertainties are graphically
represented  in Fig. \ref{hr_sc}. In the same figure, we have also
plotted the expected $H\!R1$ and $H\!R2$ values for a power-law and an
optically thin thermal plasma ({\sc mekal}) emission model with many
different values of the power-law photon index ($\gamma$), the plasma
temperature ($T$) and the intervening absorbing column ($N_{rm H}$),
computed with simulations using {\sc xspec}.

Source A is located in a region of the $H\!R1$ and $H\!R2$ plane which
correspond to very hard spectra. In particular, the non-thermal power
law seems to be preferred over the thermal plasma model, since for Source
A to be of thermal origin its temperature should be $>10^8-10^9$ K, which
seems unreasonable high. The $N_{rm H}$ of Source A is poorly constrained by
this analysis, since we have only an upper limit between $10^{22}$ and
$10^{23}$ cm$^{-2}$. We recall however, that MECS is not the best
suited instrument to measure low values of $N_{rm H}$ accurately, because
its effective area drops quickly to zero below 1 keV. This is shown in
Fig \ref{hr_sc} by the proximity of the symbols corresponding to
$10^{20}$ and $10^{21}$ cm$^{-2}$. As for Source B, the associated
hardness ratios show that it is significantly softer than Source A. It
could be a non-thermal source with $\gamma\sim 3$, or a thermal source
with a temperature of a few $10^7$ K. In any case, it seems to be very
absorbed, with $N_{rm H}>10^{22}$ cm$^{-2}$.

We have also attempted to extract the spectrum of Source A, using the
same extraction region and background used for hardness ratios. We have
used the LECS even in absence of Source A detection in order to provide
a more stringent lower limit to the value of interstellar absorption.
In fact, as the hardness ratio scatter plot suggests, MECS data yields only
an upper limit to the $N_{rm H}$, while the inclusion of LECS yields
also a lower limit to this parameter.  The spectra were rebinned to
oversample the FWHM of the energy resolution by a factor 3 and to have
additionally a minimum of 30 counts per bin to allow use of the
$\chi^2$ statistic.  Data were selected in the energy range 0.1--4~keV
for the LECS and 1.8--10~keV for the MECS.  The number of LECS and MECS
counts in the spectra after background subtraction is 20 and 307.  The
photo-electric absorption cross sections of \cite*{mm83} and the solar
abundances of \cite*{ag89} are used throughout.

\begin{figure}
  \centerline{\psfig{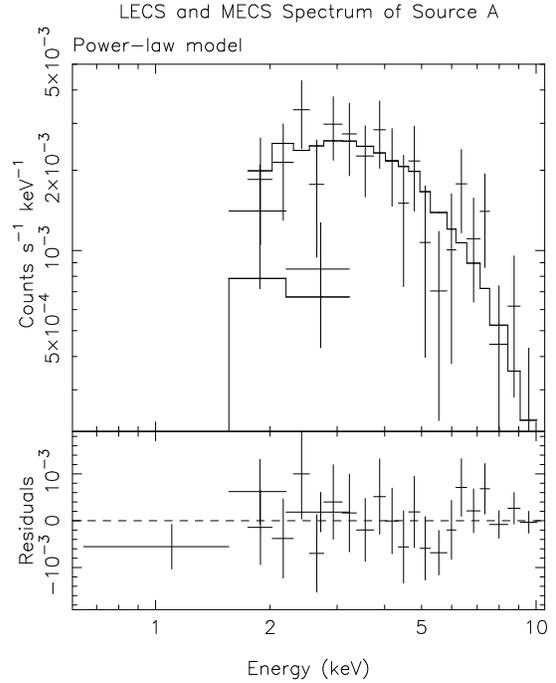}}
  \caption{\sax\  LECS and MECS spectrum of Source A. The best-fit power-law
  model is also shown together with the residuals in 
  counts~s$^{-1}$~keV$^{-1}$}
  \label{pow}
\end{figure}

Absorbed optically thin thermal plasma ({\sc mekal} in {\sc xspec}),
power-law, thermal bremsstrahlung and black-body models were tried.  At
first, we used a free normalization constant between LECS and MECS.
Since the fitted values of the constant were between 0.7 and 2.0, we
fixed its value to 1.  All models provide an adequate description of
the spectrum.  In agreement to the findings obtained with hardness ratios, 
the best-fit temperature of the {\sc mekal} fit
is very high and indicates that the power-law model may provide a more
reasonable representation of the spectrum. Fig.~\ref{pow} shows the
observed count spectrum together with the best-fit power-law model, and
Table \ref{fit} summarizes the fit results. We have verified that the
HPGSPC count-rate predicted by the best-fit models is a factor
of 5 below the upper limit given in Sect. 2.


\begin{table*}
\caption[]{Best-fit spectral parameters for the 2 strongest X-ray sources 
within \src. The flux is unabsorbed in the 2--10 keV energy range for
\sax\  and 0.1--2.4 keV energy range for \ros\  in units of
$10^{-12}$~erg cm$^{-2}$ s$^{-1}$. For the {\sc mekal} fits the abundances
were fixed at solar values. $\gamma$ is the power-law photon index
and kT is temperature. dof is the number of degrees of freedom.}
\begin{tabular}{lcccc} 
\hline\noalign{\smallskip}
Model & Parameter & ${\rm N_{rm H}}$ & Flux$\times 10^{-12}$   & $\chi^2$/dof \\
      &           & (cm$^{-2}$) & (erg cm$^{-2}$ s$^{-1}$) &  \\ 
\noalign{\smallskip\hrule\smallskip}
  & \multicolumn{4}{c}{Source A (LECS and MECS)} \\
Power-law & $\gamma = 1.8 ^{+0.8}_{-0.7}$ & $1.4^{+3.5}_{-0.6}\times 10^{22}$ & 
 $1.2^{+0.8}_{-0.5}$ & 13/20 \\
{\sc mekal} & kT$>2.6$ & $1.7^{+2.9}_{-1.0}\times 10^{22}$
& $1.2_{-0.2}^{+1.8}$ & 12/20 \\
{\sc Bremss} & kT$>4.7$ & $1.7^{+1.3}_{-0.6}\times 10^{22}$
& $1.2_{-0.2}^{+0.8}$ & 13/20 \\
Black-body & kT$=1.5^{+0.3}_{-0.3}$ & $<1.5\times 10^{22}$
& $0.9_{-0.2}^{+0.3}$ & 18/20 \\ \hline
  & \multicolumn{4}{c}{Source C (PSPC)} \\
Power-law & $\gamma = 2.1 ^{+0.5}_{-0.5}$ & $1.6 ^{+1.5}_{-1.1}\times 10^{20}$ & 
 $2.0_{-0.3}^{+0.3}$ & 23/17 \\
{\sc mekal} & kT=$2.6 ^{+3.4}_{-0.6}$ & $<1.5\times 10^{19}$
& $1.2_{-0.2}^{+0.3}$ & 26/17 \\
{\sc Bremss} & kT$=1.1^{+1.0}_{-0.4}$ & $1.0^{+0.7}_{-0.4}\times 10^{20}$
& $1.6_{-0.5}^{+0.9}$ & 19/17 \\
Black-body & kT$=0.18^{+0.02}_{-0.03}$ & $<1.5\times 10^{19}$
& $1.2_{-0.1}^{+0.1}$ & 27/17 \\
\noalign{\smallskip}
\hline
\end{tabular}
\label{fit}
\end{table*}

For the Source C, we extracted the spectrum from a circular region of
$5^\prime$ radius, which corresponds to the 90\% encircled energy
radius of the PSPC PSF at $50^\prime$ off-axis. The background was
extracted from an annulus between $5\arcmin$ and $8\arcmin$. The
spectrum was rebinned into 10 channels to oversample the energy
resolution of the instrument and was fit with the same emission models
as before using the appropriate response matrix and the ancillary file
created with the task {\sc pcarf} of FTOOLS 5.0. The background
subtracted spectrum has 466 counts. The fit results are summarized in
Table~\ref{fit}. Both thermal and non-thermal models are formally
acceptable, but the {\sc mekal} model systematically underestimates the
flux below 0.8 keV, while the Bremsstrahlung model provide the best
$\chi^2$ value.


\section{Discussion}

The X-rays emission we have detected in the direction of \src\  is not
well correlated with the radio shell: essentially, we have failed to
reveal the thermal X-ray emission which should trace the interaction
between the remnant and the interstellar medium. We have already noted
that we see an enhancement at $14^{\rm h}53^{\rm m}40^{\rm s}$
$-58\degr 48\arcmin$ which is located almost on top of the radio
filament, but further and deeper observations are needed to confirm
this detection. On the other hand, we have detected three X-ray sources
in the direction of the SNR \src, which are rather compact and loosely
correlated with the position of the radio shell.


We have investigated whether the X-ray emission of Sources~A and C
could be due to stars within the uncertainty regions.  We extracted
lists of stars from the USNO-A2.0 catalog (\cite{mon98}) within
$2^\prime$ from Sources A and C, and considered the brightest star in
each list. For Source A, there is an object with R$=13.9$ and B$=16.9$
which implies an M2 main sequence star. However, using the relation
$N_{rm H} = 1.79 \times 10^{21} \times 3.1 \times E_{B-V}$ (\cite{ps95})
with the observed $N_{rm H} \sim 1.5 \times 10^{22}$ cm$^{-2}$ (Table
\ref{fit}), we estimate a $E(B-V) \sim 2.5$, and the correction to
$B-R$ is certainly as large. Therefore, the star has an unabsorbed
$B-R\sim 0.5$, which correspond to an F star.  Since the unabsorbed V
magnitude should be around 14 and the unabsorbed 0.3--3.5~keV flux is
0.8--2.3$\times 10^{-12}$ erg cm$^{-2}$ s$^{-1}$, we calculate that the
ratio of X-ray to optical flux ($f_{\rm X} / f_{\rm V}$) is at least a
factor of 10 higher than expected for a F star (see e.g.,
\cite{mgw88}).  Similarly, from the observed $E(B-V)$ of the star
GSC08693-00036 in the uncertainty region of Source~C, we infer a K0
main sequence star at 90~pc. This is in agreement with the parallax
measured by Hipparcos (\cite{esa97}), which yields a lower limit of 11
pc to the distance to GSC08693-00036.  The observed 0.3--3.5~keV flux
is $1-3\times 10^{-12}$ erg cm$^{-2}$ s$^{-1}$, therefore, following
\cite*{mgw88}, its $f_{\rm X} /f_{\rm V}$ is consistent with a K star.
The plasma temperature derived by the fit is in agreement with the
temperature expected in a stellar corona.  Thus, while Source~C may be
identified with emission from a foreground star, Source~A appears
unlikely to be so.



Source B is a very interesting object since it is probably the X-ray
counterpart of the IRAS14498-5856 UC H~{\sc ii} region and, to our
knowledge, no other similar cases have been yet reported in the
literature.  \cite*{whr97}, using the methanol maser velocity, reported
a kinematic distance to IRAS14498-5856 of 4.3 or 11.1~kpc (depending on
whether the source is on the near, or the far side, of the Galaxy).
Using a thermal plasma emission model with a temperature of few $\times
10^7$ K (the model compatible with the hardness ratio analysis, Fig.
\ref{hr_sc}), the MECS Source~B has a flux of $7.2\pm 1.4\times
10^{-13}$~erg~cm$^{-2}$~s$^{-1}$, corresponding to $1.5\pm0.3\times
10^{33}$~erg~s$^{-1}$ at a distance of 4.3 kpc and $\sim 10^{34}$
erg~s$^{-1}$ at a distance of 11.3 kpc. The former luminosity is high
even for a young intermediate (2--5 M$_\odot$) or high mass ($>5$
M$_\odot$) star, but flares can increase luminosity. \cite*{onu99}
detected a quiescent X-ray emission of $10^{32}$ erg s$^{-1}$ from
SSV63, a 2-5 M$_\odot$ object in the dark cloud L1630, and a flare of
$10^{33}$ erg s$^{-1}$. \cite*{htb00} detected a flare from a more
massive Herbig Be star of $5\times 10^{32}$ erg s$^{-1}$. In all the
cases, the measured temperature are between 1 and 10 keV, in agreement
with the one inferred for Source B, and the flare timescale is larger
than $10^4$ s. Therefore, it may be the case that we have detected a
flaring young massive star, even if the light-curve we have extracted in our
26 ksec observation do not show particular evidence of variability. The
large absorption we measure toward Source B is also in agreement with
the explanation in terms of an embedded massive star, since the typical
value for this kind of objects is above $10^{22}$ cm$^{-2}$.
If Source B is really the hard X-ray counterpart of the young massive
star of the UC H~{\sc ii} region, then we favor the lowest distance,
because at 11 kpc its luminosity would be higher than the one expected
even in case of flare.  

Alternatively, a population of low mass pre-main sequence stars can also
produce hard X-rays. For instance, \cite*{hc97} propose this last
explanation for the hard X-ray emission from the H~{\sc ii} region W3
observed with ASCA with luminosity of $2.6\times 10^{33}$~erg~s$^{-1}$
in the 2--10 keV energy range, similar to the luminosity of Source B.
However, since the typical X-ray luminosity of low mass protostars are
in the range $10^{30}$ to $10^{31}$~erg~s$^{-1}$, a large number of
objects are required inside the \sax\  positional error circle of
$2^\prime$ (2.5 pc at 4.3 kpc) to account for the luminosity of our
source. Unfortunatly, it is not clear if IRAS14498-5856 is embedded in
a larger star forming region.  Further radio and X-ray observations are
needed to study the environment of IRAS14498-5856 and to better locate
Source B, and therefore remove this ambiguity.

An independent hint about the nature of Source B comes from its IR flux.
Using the flux densities at 60 and 100 $\mu$m of 1133 and 2208 Jy listed in
the IRAS Point Source Catalog for this source and the relation F$_{\rm
ir} = 1.3\times 10^{-11} (2.58 f_{60}+f_{100})$~erg~cm$^{-2}$ s$^{-1}$,
adapted from \cite*{rls88}, we derive a far infrared luminosity of
$1.4\times 10^{38}$ erg s$^{-1}$, and an F$_{\rm ir}$ /F$_{\rm x}$ well
above 100, in agreement with the value expected from a star forming
region (\cite{vpb97}). 


The association between SNRs and UC H~{\sc ii} regions may be very
common and, if established, supports a Type II SN progenitor
(\cite{bh87}).  If \src\  is associated with IRAS14498-5856, then its
diameter is 50~pc, and using the revised ${\rm \Sigma- D}$ relation
given by \cite*{cb98} we estimate an expected radio surface brightness
of 1--2$\times 10^{-21}$~W~m$^{-2}$~Hz$^{-1}$~sr$^{-1}$ at 1~GHz.  The
observed surface brightness is $3.7\times
10^{-22}$~W~m$^{-2}$~Hz$^{-1}$~sr$^{-1}$ and therefore this SNR is
underluminous with respect to the ${\rm \Sigma -D}$ relation. This may
imply an expansion into a low density medium, which in turn implies a
strongly reduced contribution of the thermal X-ray emission due to the
main shock, and a higher probability of observing non-thermal X-ray
emission due to accelerated electrons. This may explain why we do not
see extended thermal emission along the shell.  If Source~A indeed has
a hard non-thermal spectrum, it may be site of non-thermal electron
acceleration.  Similar hard point sources have been found inside IC443
and are interpreted as evidence of slow shocks propagating into
molecular clouds (\cite{bb00}).  This scenario is very probable since
\cite*{gm93} and \cite*{gfg97} have observed 1720 GHz line emission
inside this remnant at several locations, suggesting an OH maser due to
shock excitation of neutral gas of molecular clouds.  
We strongly suggest further deep X-ray observations of this
shell, in order to confirm the non-thermal nature of the emission,
detect any diffuse emission at other locations in the shell, and to
study the X-ray properties of the UC H~{\sc ii} region coincident with
Source~B.


The interstellar absorption provides additional information about the
distance. The absorption toward Source~A is $0.8-4.9\times 10^{22}$
cm$^{-2}$ (Table \ref{fit}), which implies a distance of
$2-14/\overline{n_0}$ kpc, where $\overline{n_0}$ is the mean
interstellar density in atoms cm$^{-3}$. If the mean density is $\sim
1$ cm$^{-3}$, than the distance estimated in this way is in agreement
with an association between Source A, Source B, the UC H~{\sc ii} region
and the SNR. On the other hand, the upper limit to the N$_{\rm H}$ is
greater than the galactic absorption in this direction.  As for
Source~C the low N$_{\rm H}$ implies a distance of $<$1 kpc, which is
consistent with the association with the K0 star located in the local
cavity at $d<100$ pc studied by \cite*{slc99}.

\begin{acknowledgements}

\sax\  is a joint Italian-Dutch program.  F.~Bocchino acknowledges
an ESA Research Fellowship.  The MOST is operated by the University of
Sydney with support from the Australian Research Council and the
Science Foundation for Physics within the University of Sydney. This
research has made use of data obtained through the High Energy
Astrophysics Science Archive Research Center Online Service, provided
by the NASA/Goddard Space Flight Center, and of data obtained through
SIMBAD, provided by the Centre de Donn\'ees astronomiques de
Strasbourg.

\end{acknowledgements}

\bibliography{references}
\bibliographystyle{aabib}

\end{document}